\documentclass[aps,prl,twocolumn,superscriptaddress, 10pt, nobibnotes, floatfix]{revtex4-2}
\usepackage{amsmath,amssymb}
\usepackage{graphicx}
\usepackage{calc,accents}
\usepackage{xcolor}
\usepackage{comment}

\usepackage[normalem]{ulem}

\usepackage[colorlinks=true, linkcolor=magenta, urlcolor=teal, citecolor=teal, filecolor=magenta]{hyperref}
\usepackage{orcidlink}

\newcommand{\dfracp}[2]{\dfrac{\partial #1}{\partial #2}}
\newcommand{\ave}[1]{\left\langle #1 \right\rangle}
\newcommand{\aveini}[1]{\left\langle #1 \right\rangle}
\newcommand{\avetau}[1]{\left\langle #1 \right\rangle_{\rm \tau}}
\newcommand{\avenoise}[1]{\left\langle #1 \right\rangle}
\newcommand{\wt}{\widetilde}
\newcommand{\wh}{\widehat}
\newcommand{\wtt}[1]{\accentset{\approx}{#1}}

\begin{document}


\title{Fluctuation--Response Relation in Finite-Size Noisy Coupled Phase Oscillators}
\author{Mrinal Sarkar\,\orcidlink{0000-0003-3112-6530}}
\email{Corresponding author; email: sarkar@thphys.uni-heidelberg.de}
\affiliation{Institut für Theoretische Physik, Universität Heidelberg, 69120 Heidelberg, Germany 
}
\author{Yoshiyuki Y. Yamaguchi\,\orcidlink{https://orcid.org/0000-0002-8825-6718}}
\email{yyama@amp.i.kyoto-u.ac.jp}
\affiliation{Graduate School of Informatics, Kyoto University, Kyoto 606-8501, Japan}

\begin{abstract}
  Fluctuation--response relations (FRRs) provide a fundamental relation
  between spontaneous fluctuations and the linear response to external perturbations,
  yet their validity in non-equilibrium systems remains an open problem.
  In a general class of noisy, finite-size, mean-field models of coupled phase oscillators,
  we reveal two types of FRRs in the incoherent phase below the synchronization transition.
  The first, corresponding to the infinite-size limit (I-FRR), involves the convolution between the external force
  and the correlation function in the time domain.
  The second, the finite-size counterpart (F-FRR),
  is governed by finite-size fluctuations and carries a
  correction factor: a spectrum function whose roots correspond to eigenvalues, or Landau poles, of the linear operator.
  The finite-size fluctuations are suppressed in the weak-coupling or strong-noise limits, and F-FRR reduces to I-FRR.
  Our analytical findings are verified by extensive numerical simulations.

\end{abstract}

\maketitle

{\it Introduction---} 
Fluctuations in stable stationary states play a key role in both equilibrium and nonequilibrium statistical physics\,\cite{Evans1993,Kurchan1998,Jarzynski2000,Sevick2008}, from driving phase transitions to relaxation and transport. For a single Brownian particle in equilibrium, its fluctuations are related to the linear response to an external perturbation, known as the fluctuation-response relation (FRR) since the early 20th century\,\cite{Einstein1905,Nyquist1928,Onsager1931,Callen1951}. For systems driven out of equilibrium, however, this relation is no longer straightforward. In the absence of any general framework connecting these two,
significant efforts have been devoted to understanding various aspects of FRRs in certain classes of nonequilibrium systems\,\cite{Agarwal1972,Baiesi2009,Prost2009,Seifert2010,Baiesi2013,Altaner2016,Chun2021,Jung2021,Shiraishi2023,Aslyamov2025,Evans2002,Evans2005,Searles2007,Dechant2020,Kwon2026,Cugliandolo1997,Deco2023,Monti2025}, see\,\cite{Kubo1966,Marconi2008,GomesFilho2025} for reviews. Most of these studies consider either a single particle or an interacting system in the thermodynamic limit. However, we ask: what happens when the system is not only interacting and driven out of equilibrium, but also finite?

Beyond its fundamental importance, experiments and simulations necessarily involve finite systems, making finite-size fluctuations unavoidable. Such fluctuations can be probed experimentally using scattering techniques\,\cite{VanHove1954, Squires1996, BernePecora2000, HansenMcDonald2013}. However, direct measurement becomes challenging at low temperatures or small scattering angles. In such cases, an FRR offers an alternative yet powerful tool to probe the underlying correlations.

\begin{figure}[htbp]
  \centering
  \includegraphics[width=0.9\linewidth]{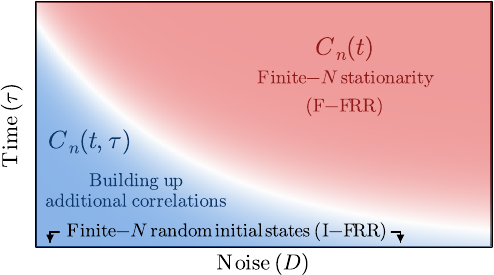}
  \caption{Schematic crossover in the $D$ (noise) -  $\tau$ (time) plane within the nonsynchronized phase. As $\tau$ progresses, the transient correlation function $C_n(t,\tau)$ (blue), for mode $n$, crosses over to the time-averaged, stationary correlation function $C_n(t)$ (red); see the main text for their definitions. Additional correlations are generated by finite-size effects while relaxing to the finite-$N$ stationary state,
    whereas $C_{n}(t)$ approaches $C_n^{\rm ini}(t) := C_n(t,\tau=0)$ as $D$ increases.
    $C_n(t)$ satisfies the finite-size fluctuation-response relation (F-FRR), while the infinite-size FRR (I-FRR) is satisfied only by $C_n^{\rm ini}(t)$.
  }
  \label{fig:Schematic}
\end{figure}

We address this issue of fluctuations and response in a general class of globally coupled phase oscillator systems, which describe macroscopic rhythmic phenomena and serve as a critical system at the interface of nonlinear dynamics and statistical physics\,\cite{Acebron2005, PikovskyRosenblum2015}. In finite-$N$ systems, the source of fluctuations is {\it twofold}: {\it external noise} and {\it finite size}. Finite-size fluctuations in coupled oscillator systems have recently attracted renewed attention, both for the phenomena they generate\,\cite{Snyder2021,Yue2024,Suman2024,Irvine2026} and for the analytical challenges they pose\,\cite{Yoon2015,Tyulkina2018,Buendia2025}. Progress has come through finite-size scaling\,\cite{Daido1990,Hong2007,Lee2014,Hong2015,Coletta2017,Park2024}, kinetic theory\,\cite{Hildebrand2007,BuiceChow2007}, collective co-ordinate approach\,\cite{GottWald2017}, microscopic correlations\,\cite{Das2018}, power spectrum\,\cite{Kati2024}, shot-noise approach\,\cite{Kirillov2025}, and very recently mean-field theory\,\cite{Omelchenko2026}. On the response side, susceptibilities and related response functions have been studied separately\,\cite{Sakaguchi1988,Daido2015,Terada2020,Amadori2022}.

Despite these developments, an FRR for macroscopic observables, such as the order parameter, remains elusive in finite-size coupled phase-oscillator systems incorporating both sources of fluctuations. Establishing such a relation is of fundamental and practical importance. In this Letter, we derive two FRRs in this nonequilibrium setting for general phase-oscillator systems in the nonsynchronized phase: one is the exact relation accounting for finite-$N$ fluctuations, while the other corresponds to the thermodynamic limit $N\to\infty$ and is obtained by focusing on the initial state drawn from the infinite-size stationary state. A system starting from the infinite-size stationary state relaxes to the finite-size stationary state by generating an additional correlation for any large but finite $N$. The generated correlation is expressed by the spectrum function, which provides the stability of the infinite-size stationary state. To establish these results, we apply the Dean-Kawasaki formalism\,\cite{Dean1996,Kawasaki1998,Illien2025,Cornalba2023} to finite-size noisy coupled oscillators. Our main results are summarized in Fig.~\ref{fig:Schematic}.

\paragraph{Main results ---}
We consider a general class of $N$ globally-coupled phase oscillators. The phase $\theta_{j}$ of $j$-th ($j=1, 2, \ldots, N $) oscillator evolves as
\begin{equation}
  \dot{\theta}_j = \omega_j + \frac{1}{N}\sum_{k=1}^N \Gamma(\theta_j - \theta_k) 
  + H(\theta_j,t) + \sqrt{2D} \xi_j(t),
  \label{eq:evolution1}
\end{equation}
where $\omega_{j}$ is the natural frequency, independently and randomly drawn from the distribution $g(\omega)$, $\Gamma(\theta)$ is the interaction kernel, $H(\theta,t)$ is a weak external force, and $D>0$ is the diffusion constant. The $\xi_{j}$'s are independent Gaussian white noises with $\langle \xi_j(t)\rangle = 0$ and $\langle \xi_j(t)\xi_k(t')\rangle = \delta_{j,k}\delta(t-t')$, where $\delta_{j,k}$ is the Kronecker delta, $\delta(t)$ is the Dirac delta function, and $\ave{\cdots}$ represents the average over realizations. The initial conditions $\{\theta_{j}\}$ at $t=0$ are independently and randomly drawn from the uniform distribution on $(-\pi,\pi]$. We stress that the initial time $t=0$ is special since the initial conditions are not stationary in a finite-$N$ system.

We are interested in the fluctuation-response relation of macroscopic variables, which are the complex order parameters $Z_{n}~(n\in\mathbb{Z})$ defined by
\begin{equation}
  Z_{n}^{h}(t) = \dfrac{1}{N} \sum_{k=1}^{N}e^{in\theta_{k}(t)}.
  \label{eq:Zm}
\end{equation}
The superscript $h$ represents the presence of the external force $H\not\equiv 0$ and the order parameters with $H\equiv 0$ are simply denoted by $Z_{n}(t)$. From now on we focus on $n\neq 0$ since $Z_{0}^{h}(t)\equiv 1$ is trivial. In the nonsynchronized phase we have $\ave{Z_{n}(t)}=0$. Keeping this in mind, the response is defined by
\begin{equation}
  R_{n}(t) = \aveini{Z_{-n}^{h}(t)}.
  \label{eq:Rnt}
\end{equation}
The indexing of $R_{n}$ is chosen for later convenience. While measuring the response, the external force $H$ is switched on at $t=0$, but $R_n(t)$ depends only on the time elapsed since the kick and not on the kick-in time. In the absence of the external force ($H=0$), we define the correlation function of $Z_{n}$ as $C_{n}(t, \tau) = \avenoise{ Z_{-n}(t+\tau) \, Z_{n}(\tau)}$ which measures the correlation at a time lag $t$ when the system is observed at time $\tau$, elapsed since the initial condition. $C_{n}(t,\tau)$ allows one to track the relaxation from the initial-value correlation
\begin{equation}
  C_{n}^{\rm ini}(t) = C_{n}(t, 0),
  \label{eq:Cninit}
\end{equation}
obtained by fixing the reference time at $\tau=0$, to the stationary correlation
\begin{equation}
  C_{n}(t) = \avetau{C_{n}(t, \tau) },
  \label{eq:Cnt}
\end{equation}
obtained by averaging over $\tau$. At $t=0$, both $C_{n}^{\rm ini}(0)$ and $C_{n}(0)$ reduce to the (real-valued) variance of $Z_n$.

To derive the FRRs, we formulate the dynamics\,\eqref{eq:evolution1} in terms of the Dean--Kawasaki (DK) equation\,\cite{Majumder2025}
\begin{equation}
  \dfracp{F}{t} = - \dfracp{(V[F]F)}{\theta} + D \dfracp{{}^{2}F}{\theta^{2}}
  + \dfrac{1}{\sqrt{N}} \dfracp{}{\theta} \left[ \sqrt{2DF} \zeta(\theta,\omega,t) \right].
  \label{eq:DK}
\end{equation}
Here $F(\theta,\omega,t)$ is the probability density function, $V[F]=\omega+v[F]+H$ is the deterministic velocity field with $v[F](\theta,t)=\int_{\mathbb{R}}d\omega\int_{-\pi}^{\pi} d\theta' \Gamma(\theta-\theta') F(\theta',\omega,t)$, and $\zeta$ is Gaussian white noise satisfying $\avenoise{\zeta(\theta,\omega,t)}=0$ and  $\avenoise{\zeta(\theta,\omega,t)\zeta(\theta',\omega',t')} =\delta(\theta-\theta')\delta(\omega-\omega')\delta(t-t')$. We refer to the last term in Eq.~\eqref{eq:DK} as the DK term, which explicitly captures finite-$N$ fluctuations. In the absence of an external field ($H\to 0$) and in the limit $N\to\infty$, Eq.\, \eqref{eq:DK} admits the stationary nonsynchronized state denoted by $F^{\rm st}(\theta,\omega)=g(\omega)/(2\pi)$ below the critical coupling strength. Note that the initial conditions in the $N$-body dynamics follow $F^{\rm st}$, and $C_{n}^{\rm ini}$ corresponds to the stationary correlation in the limit $N\to\infty$.

We assume that $|H| \ll1$ and $N$ is sufficiently large with $|H|\gg 1/\sqrt{N}$, so that the response can be distinguished from the finite-$N$ fluctuations. Expanding $F=F^{\rm st}+f$, then yields the linearized DK equation
\begin{equation}
  \dfracp{f}{t}
  = -\omega\dfracp{f}{\theta} - F^{\rm st}\dfracp{(v[f]+H)}{\theta}
  + D \dfracp{{}^{2}f}{\theta^{2}}
  + \sqrt{ \dfrac{2DF^{\rm st}}{N} } \dfracp{\zeta}{\theta}.
  \label{eq:linear-DK}
\end{equation}
We solve Eq.\,\eqref{eq:linear-DK} by performing the Fourier transform in $\theta$ and the Laplace or Fourier transform in $t$, and assuming $\ave{f^{\rm ini}(\theta,\omega)\zeta(\theta',\omega',t')}=0$
for the initial state $f^{\rm ini}(\theta,\omega)=f(\theta,\omega,0)$ (see Appendix A in End Matter and \cite{SM}). We denote the $n$th Fourier component of $H(\theta,t)$ by $\wt{H}_{n}(t)$, and its Laplace transform by $\wh{H}_{n}(s)$. The same notation is used for the other functions. By using the solution to Eq.\,\eqref{eq:linear-DK} and the spectrum functions $\Lambda_{n}(s)$, we obtain the Laplace transforms of the correlation functions as
\begin{equation}
  \wh{C}_{n}^{\rm ini}(s)
  = \dfrac{1}{-inN} \dfrac{1-\Lambda_{n}(s)}{\wt{\Gamma}_{n}\Lambda_{n}(s)},
  \label{eq:correlation-ini}
\end{equation}
\begin{equation}
  \wh{C}_{n}(s)
  = \dfrac{1}{-inN\Lambda_{-n}(-s)} \dfrac{1-\Lambda_{n}(s)}{\wt{\Gamma}_{n}\Lambda_{n}(s)},
  \label{eq:correlation}
\end{equation}
and of the response as
\begin{equation}
  \wh{R}_{n}(s) = \dfrac{1-\Lambda_{n}(s)}{\wt{\Gamma}_{n}\Lambda_{n}(s)} \wh{H}_{n}(s).
  \label{eq:response}
\end{equation}
The FRRs in the Laplace domain are therefore
\begin{equation}
  \wh{R}_{n}(s) = -in N \wh{C}_{n}^{\rm ini}(s) \wh{H}_{n}(s)
  \label{eq:I-FRR}
\end{equation}
and
\begin{equation}
  \wh{R}_{n}(s) = -in N \Lambda_{-n}(-s) \wh{C}_{n}(s) \wh{H}_{n}(s).
  \label{eq:F-FRR}
\end{equation}
The relations \eqref{eq:I-FRR} and \eqref{eq:F-FRR} are the main results of this Letter. The spectrum function $\Lambda_{n}(s)$ is defined by
\begin{equation}
  \Lambda_{n}(s)
  = 1 + \wt{\Gamma}_{n} \int_{L} \dfrac{ing(\omega)}{s+in\omega+Dn^{2}} d\omega,
  \label{eq:SpectrumFunction}
\end{equation}
where $L$ is the Landau contour \cite{Landau1946}, which lies on the real axis for ${\rm Re}(s)>-Dn^{2}$ but is smoothly modified to avoid the pole at $\omega=i(s+Dn^{2})/n$ for ${\rm Re}(s)\leq -Dn^{2}$ (see Appendix B in End Matter). The roots of $\Lambda_{n}(s)$ are the eigenvalues or Landau poles of the linearized operator and determine the spectrum stability of $F^{\rm st}$\,\cite{StrogatzMirollo1991}.

The FRR \eqref{eq:I-FRR} is not affected by the DK term, because of the independence between the initial state $f^{\rm ini}$ and the noise $\zeta$. Therefore, the spectrum function $\Lambda_{-n}(-s)$ in the FRR \eqref{eq:F-FRR} provides the finite-size corrections to the FRR \eqref{eq:I-FRR}. In other words, in a finite-$N$ system, temporal evolution generates additional correlations among oscillators that are absent in the infinite-$N$ description. Based on this discussion, we refer to Eq.~\eqref{eq:I-FRR} as the infinite-size FRR ({\rm I-FRR}) and Eq.~\eqref{eq:F-FRR} as the finite-size FRR ({\rm F-FRR}).

A type of $C_{n}^{\rm ini}(t)$ is discussed theoretically in Ref.\,\cite{Omelchenko2026} in the finite-$N$ Sakaguchi-Kuramoto model\,\cite{Sakaguchi1986} with the aid of the Ott-Antonsen reduction \cite{Ott2008,Ott2009}.
We stress that the correlation $C_{n}(t)$ defined by Eq.~\eqref{eq:Cnt} is more directly relevant to experimental measurements and that the relation\,\eqref{eq:F-FRR} is nontrivial, since straightforward computations give a relation involving $\wh{C}_{\pm n}$ and $\wh{R}_{\pm n}$; see Appendix A in End Matter.

Three remarks on the FRRs \eqref{eq:I-FRR} and \eqref{eq:F-FRR} are in order. (i) The response \eqref{eq:response} is the same in the {\rm I-FRR} \eqref{eq:I-FRR} and the {\rm F-FRR} \eqref{eq:F-FRR} due to the scale separation $|H|\gg 1/\sqrt{N}$. (ii) The correlations are of $\mathcal{O}(1/N)$ since $Z_{n}=\mathcal{O}(1/\sqrt{N})$, and both sides of Eqs.~\eqref{eq:I-FRR} and \eqref{eq:F-FRR} are of $\mathcal{O}(N^{0})$. (iii) Recalling the definition \eqref{eq:SpectrumFunction}, we have $\Lambda_{-n}(-s)\to 1$ in the weak coupling limit $\wt{\Gamma}_{n}\to 0$ or the strong diffusion limit $D\to\infty$. In the former limit, interactions are too weak to generate additional correlations, whereas in the latter, a strong $D$ suppresses the finite-size fluctuations.

\begin{figure}[htbp]
  \centering
  \includegraphics[width=0.99\linewidth]{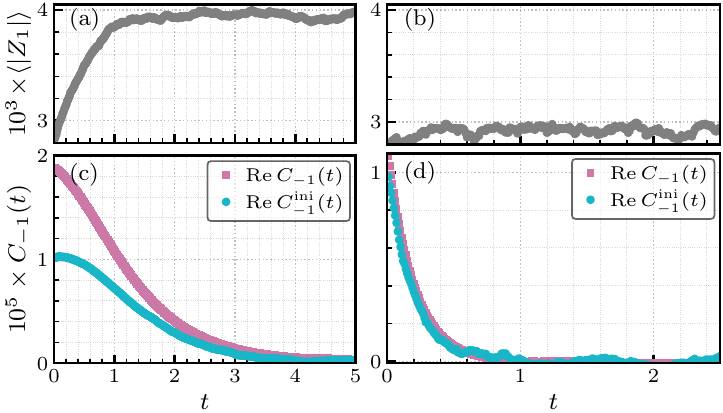}
  \caption{Order-parameter magnitude and correlation functions in the absence of external perturbation ($h=0$), for $K=1$ at weak noise $D=0.5$ [panels (a),(c)] and strong noise $D=5.5$ [panels (b),(d)]. Top row: order-parameter magnitude $\ave{|Z_1(t)|}$, relaxing from its
initial value to a finite-$N$ stationary value. Bottom row: real
part of the two-point correlation functions $C_{-1}(t)$ (pink squares) and
$C_{-1}^{\rm ini}(t)$ (cyan circles). At $D=0.5$ the two correlation
functions are clearly distinguishable over the whole observed time window, whereas
at $D=5.5$ they collapse onto a single curve.}
\label{fig:r_and_correlation}
\end{figure}

\paragraph{Numerical verification ---}As a testbed, we adopt the paradigmatic coupled phase oscillator system, the Kuramoto model\,\cite{Kuramoto1984}, with coupling
kernel $\Gamma(\theta)=-K\sin\theta$ (i.e., $\wt{\Gamma}_{\pm 1}=\pm iK/2$), subject to noise of strength $D$ and the external force $H(\theta,t) = - h \, \Theta(t)\, \sin(\theta - \omega_{\mathrm{ext}}t)$
(i.e., $\widetilde{H}_{\pm 1}(t)=\pm i (h/2) \Theta(t) e^{\mp i\omega_{\rm ext}t}$), where $\Theta(t)$ is the unit step function. The natural frequency distribution is the normal Gaussian, $g(\omega)=\exp(-\omega^{2}/2)/\sqrt{2\pi}$. The spectrum function \eqref{eq:SpectrumFunction} gives the noise-dependent critical coupling $K_{\rm c}(D) = 2\sqrt{{2}/{\pi}}\, {e^{-D^{2}/2}} / {\mathrm{erfc}(D/\sqrt{2})}$, which recovers $K_{\rm c}(0)=2\sqrt{2/\pi} =1.595769\ldots$. We are interested in the nonsynchronized regime $K<K_{\rm c}(D)$.

We perform extensive $N$-body stochastic simulations with $N=10^{5}$, using the Euler-Maruyama algorithm with a timestep $\Delta t=0.01$. The correlation functions $C_{-1}(t)$ and $C_{-1}^{\rm ini}(t)$ are computed
with zero external force ($h=0$), and the response $R_{-1}(t)$ with $h=0.01$ and $\omega_{\rm ext}=0.1$, at various $D$ and $K$. The $\tau$-average is taken over the interval $\tau \in [0,T-t]$, with $T=10$. The strength $h$ is chosen to dominate over finite-size fluctuations, i.e., $h\gg 1/\sqrt{N}$.  All numerical results reported here are obtained by averaging over $2 \times 10^{3}$ realizations.
 
We first examine the correlation functions of the order parameter in the absence of an external perturbation ($h=0$), at fixed coupling $K=1$ ($<K_{\rm c}$), comparing two noise strengths, $D=0.5$ (low-$D$
case) and $D=5.5$ (high-$D$ case), see Fig.~\ref{fig:r_and_correlation}. The order-parameter magnitude $\ave{|Z_1(t)|}$ [panels (a),(b)] relaxes from its initial value to a finite-$N$ stationary value. A remarkable difference is that $\ave{|Z_1(t)|}$ increases over time before saturating around a mean stationary value in panel (a), while it remains essentially unchanged from its initial value in panel (b). The bottom row [(c),(d)] compares the two correlation functions $C_{-1}(t)$ and $C_{-1}^{\rm ini}(t)$: at low noise ($D=0.5$) the two are clearly distinguishable throughout the observed time window, whereas at high noise ($D=5.5$) they collapse onto a single curve. This indicates that an additional correlation is generated in the low-$D$ case, in a system with a large but finite $N$, even though $K$ is well below the critical point $K_{\rm c}$.

\begin{figure}[htbp]
  \centering
  \includegraphics[width=0.99\linewidth]{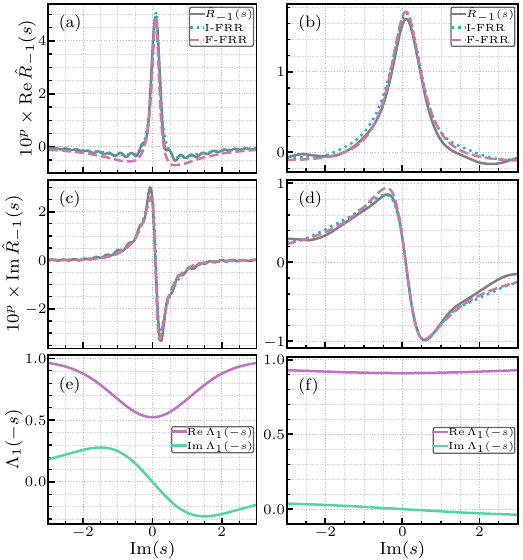}
  \caption{Verification of fluctuation-response relations (FRRs) in the Laplace domain. Panels (a-d) show the measured response spectrum $\hat{R}_{-1}(s)$ (grey) with the right-hand side of I-FRR [Eq.~\eqref{eq:I-FRR}, cyan dotted] and of F-FRR [Eq.~\eqref{eq:F-FRR}, pink dashed] predictions, evaluated along $s=\mathrm{Re} (s)+i\,\mathrm{Im}(s)$ with $\mathrm{Re} (s)=0.125$, at weak noise $D=0.5$ [panels (a),(c)] and strong noise $D=5.5$ [panels (b),(d)], for $K=1$. Top row: $\mathrm{Re}\,\hat{R}_{-1}(s)$; middle row: $\mathrm{Im}\,\hat{R}_{-1}(s)$.
    Panels (e, f): real (purple) and imaginary (green) parts of the spectrum function $\Lambda_1(-s)$
    [Eq.~\eqref{eq:SpectrumFunction}]. For clarity, the responses are shown as $10^{p} \times \hat{R}_{-1}(s)$ with $p=2$ in panels (a,c) and $p=3$ in panels (b,d).
  } 
\label{fig:laplace_domain}
\end{figure}

\begin{figure}[htbp]
  \centering
     \includegraphics[width=0.99\linewidth]{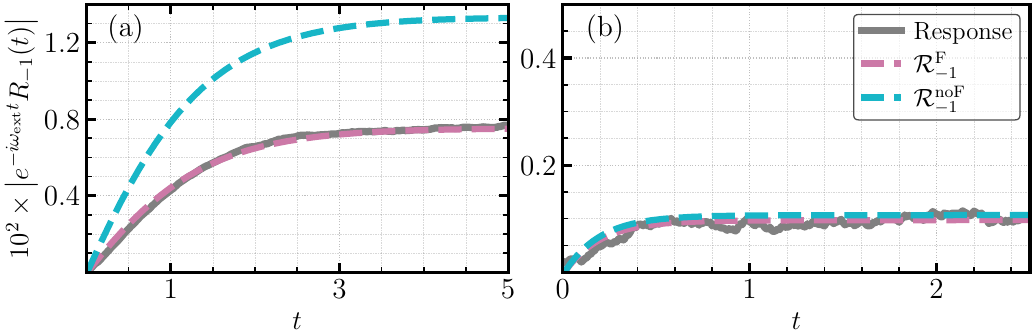}
   \caption{Verification of F-FRR in the time domain. The direct measurement of $|e^{-i\omega_{\rm ext}t}R_{-1}(t)|$ (grey) is compared with the F-FRR prediction
     $\mathcal{R}_{-1}^{\rm F}(t)$ (pink dashed), and also with $\mathcal{R}_{-1}^{\rm noF}(t)$ (cyan dashed) that neglects the finite-size correction. Here,
     $\omega_{\rm ext}=0.1$. The parameters are identical to those in Fig.~\ref{fig:laplace_domain}:
     (a) $D=0.5$ and (b) $D=5.5$ with $K=1$.
   }
  \label{fig:fig2_timedomain}
\end{figure}


\begin{figure}[htbp]
  \centering
  \includegraphics[width=0.99\linewidth]{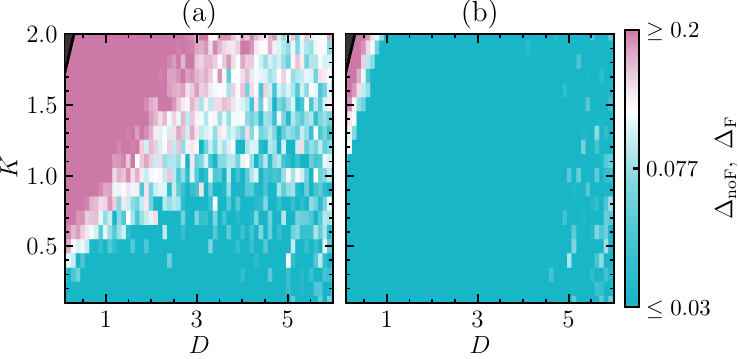}
   \caption{Validity of the F-FRR across the $(D,K)$ plane via the normalized time-domain residual $\Delta$.
     Color encodes, on a log scale, (a) $\Delta_{\rm noF}$ and (b) $\Delta_{\rm F}$:
     $\Delta\leq 0.1$ (teal) indicating good agreement and $\Delta\geq 0.2$ (magenta) indicating poor agreement. The threshold is chosen to lie within the statistical fluctuations of the simulations arising from the finite number of realizations\,\cite{SM}. The observation time $T_{\rm obs}=5$ is chosen so that the correlation has decayed to zero.
     The black region marks the synchronized phase with $K>K_{\rm c}(D)$
     and lies outside the domain of the theory.}
\label{fig:dk_validity_map} 
\end{figure}

Figure~\ref{fig:laplace_domain} presents the quantitative test of the two FRRs, when the external perturbation is switched on. We compare the measured spectrum $\hat{R}_{-1}(s)$ to the I-FRR prediction of Eq.~\eqref{eq:I-FRR} and the F-FRR prediction of Eq.~\eqref{eq:F-FRR}, both in the real and imaginary parts, at weak noise ($D=0.5$) and strong noise ($D=5.5$), together with the analytic spectrum function $\Lambda_1(-s)$ that encodes the finite-size correction [Eq.~\eqref{eq:SpectrumFunction}]. At $D=0.5$, the I-FRR prediction shows very good agreement with the measured response when $C_{-1}^{\rm ini}$ is considered directly. Also, for $C_{-1}$, the F-FRR prediction reproduces the measured spectrum essentially exactly when $\Lambda_1$ is taken into consideration, in both real and imaginary parts. At $D=5.5$, the two predictions also become indistinguishable from each other and from the measurement.

In experiments the accessible correlation function is $C_{-1}(t)$, while $C_{-1}^{\rm ini}(t)$ is a special one that does not account for correlations subsequently generated. We therefore focus hereafter on the F-FRR in time domain \cite{SM}, which reads as
\begin{equation}
  e^{-i\omega_{\rm ext}t} R_{-1}(t)
  = \dfrac{h}{2} N \Lambda_{1}(-i\omega_{\rm ext})
  \int_{0}^{t} e^{-i\omega_{\rm ext}t'}\, C_{-1}(t')\, dt',
  \label{eq:F-FRR-example}
\end{equation}
and verify its validity in the $(D,K)$ plane, thereby assessing the importance of the $\Lambda_{1}$ correction. To this end, we compare the right-hand side of Eq.~\eqref{eq:F-FRR-example}, denoted by $\mathcal{R}_{-1}^{\rm F}(t)$, against $\mathcal{R}_{-1}^{\rm noF}(t)$ defined by replacing $\Lambda_{1}(-i\omega_{\rm ext})$ with $1$. First, the comparison is performed in Fig.~\ref{fig:fig2_timedomain} at the same parameter points as in Fig.~\ref{fig:laplace_domain}.

For $K=1.0$, at weak noise $D=0.5$, the measured response is in excellent agreement with $\mathcal{R}_{-1}^{\rm F}(t)$, whereas $\mathcal{R}_{-1}^{\rm noF}(t)$ overestimates the amplitude substantially, demonstrating that the finite-size correction is not a small perturbative effect. At strong noise $D=5.5$, both predictions agree with the measurement, consistent with the expected convergence of $C_{-1}^{\rm ini}$ and $C_{-1}$ in the limit $\Lambda_{1}\to 1$.

Next, we quantify the agreement over the $(D, K)$ plane using the normalized residual
\begin{equation}
  \Delta_{\rm F}
  = \dfrac{\int_{0}^{T_{\rm obs}} | e^{-i\omega_{\rm ext}}R_{-1}(t)-\mathcal{R}_{-1}^{\rm F}(t)|^{2} dt}
  {\int_{0}^{T_{\rm obs}} | e^{-i\omega_{\rm ext}}R_{-1}(t)|^{2} dt}
  \label{eq:Delta}
\end{equation}
and $\Delta_{\rm noF}$ defined by replacing $\mathcal{R}_{-1}^{\rm F}(t)$ with $\mathcal{R}_{-1}^{\rm noF}(t)$. Figure \ref{fig:dk_validity_map}(a) shows that $\Delta_{\rm noF}$ is small only in the lower-right region,
where the finite-size correction is not essential, i.e., $\Lambda_{1}\simeq 1$. In contrast, $\Delta_{\rm F}$ remains small, within the statistical resolution of the simulations\,\cite{SM}, across nearly the entire $(D,K)$ plane,
as shown in Fig.~\ref{fig:dk_validity_map}(b), except in a neighborhood of the critical line $K_{\rm c}(D)$, indicated by the boundary of the black synchronized region, where the linear response diverges on the boundary. This demonstrates that the F-FRR \eqref{eq:F-FRR}, which captures finite-size corrections, remains valid throughout the nonsynchronized phase.


\paragraph{\it Summary ---}
To summarize, using the DK equation, we have derived theoretically the finite-size fluctuation-response relation (F-FRR) for finite-size, noisy, coupled phase-oscillator systems and confirmed its validity through systematic numerical simulations. The relation requires a dynamical correction, given by the spectrum function whose roots are the eigenvalues of the linearized operator, capturing finite-size effects. Nevertheless, the infinite-size fluctuation-response relation (I-FRR) is recovered either in the limits of weak coupling or strong diffusion, where the correction term approaches unity, or can also be recovered exactly by measuring the correlation between the initial state and its evolved counterpart.

A natural extension of the current study is to explore FRR in more complex coupling topologies, including sparse, modular, or long-range interaction graphs, which may exhibit richer finite-size effects\,\cite{Millan2020, SarkarGupta2020, SarkarEnssDefenu2024, Manik2017, ZhangTimme2023}. Phase oscillators on small-world networks, which can be viewed as a noisy Kuramoto model\,\cite{Yoneda2020}, offer a natural setting. A recent numerical study of the Kuramoto model on real neural connectomes examined microscopic FRR in a sparse, finite-dimensional network topology\,\cite{Odor2025}, in stark contrast to the all-to-all coupled, infinite-dimensional setting considered here. This further motivates a systematic study of FRR on networks with finite spectral dimensions\,\cite{SarkarEnssDefenu2024}. Another key direction is to extend the present analysis to the synchronized regime. Finally, recent experimental progress in realizing synchronization in oscillator systems, where noise and finite-size effects are naturally present, provides a promising platform to test our theoretical predictions\,\cite{Aravind2024, Freitas2026}.

\begin{acknowledgments}
M.S. acknowledges funding from the Deutsche Forschungsgemeinschaft (DFG, German Research Foundation) under Germany's Excellence Strategy EXC2181/1-390900948 (the Heidelberg STRUCTURES Excellence Cluster) and support by the state of Baden-Württemberg through bwHPC. M.S. also acknowledges helpful discussions with Shamik Gupta. Y.Y.Y. acknowledges the support of JSPS KAKENHI Grant No. JP21K03402.
\end{acknowledgments}

\clearpage
\appendix
\section*{End Matter}

\subsection{Appendix A: Derivation of the fluctuation-response relations}
\label{app_A}


We expand $f(\theta,\omega,t)$ in the Fourier series as
\begin{equation}
  f(\theta,\omega,t) = \sum_{n\in\mathbb{Z}} e^{in\theta} \wt{f}_{n}(\omega,t),
\end{equation}
and denote the Laplace transform of $\wt{f}_{n}(\omega,t)$ by $\wh{f}_{n}(\omega,s)$.
The same notation is used for the other functions.
The linearized DK equation \eqref{eq:linear-DK} is transformed to an algebraic equation,
and the solution satisfies the self-consistent equation
\begin{equation}
  \begin{split}
    & \wh{f}_{n}(\omega,s)
    = \dfrac{-inF^{\rm st}(\omega)}{s+in\omega+n^{2}D}
      \left[ \wt{\Gamma}_{n} \wh{Z}_{-n}^{h}(s) + \wh{H}_{n}(s) \right] \\
    & + \dfrac{\wt{f}_{n}^{\rm ini}(\omega)}{s+in\omega+n^{2}D}
  + \sqrt{\dfrac{2D}{N}} \dfrac{in\sqrt{F^{\rm st}(\omega)}}{s+in\omega+n^{2}D} \wh{\zeta}_{n}(\omega,s),
  \end{split}
  \label{eq:whfn}
\end{equation}
where $f^{\rm ini}(\theta,\omega)=f(\theta,\omega,0)$ and we assume
\begin{equation}
  \begin{split}
    & \aveini{f^{\rm ini}(\theta,\omega)} = 0, \\
    & \aveini{f^{\rm ini}(\theta,\omega)f^{\rm ini}(\theta',\omega')}
      = \dfrac{a}{N} \delta(\theta-\theta') \delta(\omega-\omega') F^{\rm st}(\omega), \\
    & \aveini{f^{\rm ini}(\theta,\omega)\zeta(\theta',\omega',t')} = 0.
  \end{split}
  \label{eq:average-correlation-fluctuation}
\end{equation}
A numerical check implies $a=1$.
Equation \eqref{eq:whfn} is a self-consistent equation,
since $\wh{Z}_{-n}^{h}(s)=2\pi\int_{-\infty}^{\infty} \wh{f}_{n}(\omega,s)d\omega$
includes $\wh{f}_{n}$.
Multiplying by $2\pi$ and integrating over $\omega$ in Eq.~\eqref{eq:whfn}, we have
\begin{equation}
  \begin{split}
    \wh{Z}_{-n}^{h}(s)
    & = \dfrac{\wh{H}_{n}(s)}{\Lambda_{n}(s)}
      \int_{-\infty}^{\infty} \dfrac{-ing(\omega)}{s+in\omega+n^{2}D} d\omega \\
    & + \dfrac{2\pi}{\Lambda_{n}(s)}
      \int_{-\infty}^{\infty} \dfrac{\wt{f}_{n}^{\rm ini}(\omega)}{s+in\omega+n^{2}D} d\omega \\
    & + \dfrac{1}{\Lambda_{n}(s)}
      \sqrt{\dfrac{4\pi D}{N}}
      \int_{-\infty}^{\infty} \dfrac{in\sqrt{g(\omega)}}{s+in\omega+n^{2}D} \wh{\zeta}_{n}(\omega,s) d\omega. \\
  \end{split}
  \label{eq:DK-FL-Z}
\end{equation}
The average $\aveini{\cdots}$ eliminates $\wt{f}_{n}^{\rm ini}$ and $\wt{\zeta}_{n}$,
and the Laplace transform of the response $R_{n}(t)$ is obtained as 
\begin{equation}
  \wh{R}_{n}(s) = \dfrac{1}{\wt{\Gamma}_{n}} \dfrac{1-\Lambda_{n}(s)}{\Lambda_{n}(s)} \wh{H}_{n}(s).
\end{equation}

\subsubsection{Infinite-size fluctuation-response relation (I-FRR)}
The Laplace transform of $C_{n}^{\rm ini}(t)$ is
\begin{equation}
  \wh{C}_{n}^{\rm ini}(s) = \aveini{ \wh{Z}_{-n}(s) Z_{n}(0) },
\end{equation}
where $Z_{n}(0)=\int_{-\infty}^{\infty}d\omega\int_{-\pi}^{\pi} d\theta e^{in\theta}f^{\rm ini}(\theta,\omega) =2\pi\int_{-\infty}^{\infty} \wt{f}_{-n}^{\rm ini}(\omega) d\omega$.
The average gives
\begin{equation}
  \wh{C}_{n}^{\rm ini}(s) = \dfrac{a}{-inN} \dfrac{1}{\wt{\Gamma}_{n}}
  \dfrac{1-\Lambda_{n}(s)}{\Lambda_{n}(s)}
\end{equation}
and the fluctuation-response relation \eqref{eq:I-FRR} with $a=1$.

\subsubsection{Finite-size fluctuation-response relation (F-FRR)}
The correlation $C_{n}(t)$ is computed by setting $\wh{H}_{n}\equiv 0$
and eliminating the cross terms between $\wt{f}_{n}^{\rm ini}$ and $\wh{\zeta}_{n}$
in the average $\aveini{\cdots}$.
Since $F^{\rm st}$ is stable, $\Lambda_{n}(s)$ has roots
only in the left-half plane ${\rm Re}(s)<0$.
The initial fluctuation $\wt{f}_{n}^{\rm ini}$ cannot survive in the average $\avetau{\cdots}$,
since it gives solely exponentially decaying terms.
The contribution from $\wh{\zeta}_{n}$ is, however, not trivial,
since the correlation of $\zeta(\theta,\omega,t)$ depends on $t$.
To simplify computations, instead of the Laplace transform,
we use the Fourier transform in $t$ defined by
\begin{equation}
  \wtt{\zeta}_{n}(\omega,\rho)
  = \dfrac{1}{2\pi} \int_{-\infty}^{\infty} dt e^{-i\rho t} \zeta(\omega,t),
\end{equation}
which gives
\begin{equation}
  \aveini{\wtt{\zeta}_{n}(\omega,\rho)\wtt{\zeta}_{-n}(\omega',\rho')}
  = \dfrac{1}{(2\pi)^{2}} \delta(\omega-\omega') \delta(\rho+\rho').
\end{equation}
The Fourier transform of $Z_{-n}(t)$ from $\zeta_{n}$ is
\begin{equation}
  \wtt{Z}_{-n}(\rho)
  = \dfrac{1}{\Lambda_{n}(i\rho)}
  \sqrt{\dfrac{4\pi D}{N}}
  \int_{-\infty}^{\infty} \dfrac{in\sqrt{g(\omega)}}{i\rho+in\omega+n^{2}D}
  \wtt{\zeta}_{n}(\omega,\rho) d\omega.
\end{equation}
Substituting $\wtt{Z}_{-n}(\rho)$ into Eq.~\eqref{eq:Cnt},
the Fourier transform of $C_{n}(t)$ is obtained as
\begin{equation}
  \begin{split}
    \wtt{C}_{n}(\rho)
    & = \dfrac{-1}{2\pi N} 
      \dfrac{1}{\Lambda_{n}(i\rho)\Lambda_{-n}(-i\rho)} \\
    & \times \left[
      \dfrac{1-\Lambda_{n}(i\rho)}{in\wt{\Gamma}_{n}}
      + \dfrac{1-\Lambda_{-n}(-i\rho)}{-in\wt{\Gamma}_{-n}}
      \right]. \\
  \end{split}
  \label{eq:wttCnrho}
\end{equation}
The right-hand side contains $\wh{R}_{n}(i\rho)$ and $\wh{R}_{-n}(-i\rho)$.
We will obtain the fluctuation-response relation \eqref{eq:F-FRR}
in the following three steps.

First, the definition of the Fourier transform yields
\begin{equation}
  \begin{split}
    \wtt{C}_{n}(\rho)
    & = \dfrac{1}{2\pi} \left[
      \int_{0}^{\infty} dt~ e^{-i\rho t} C_{n}(t)
      + \int_{0}^{\infty} dt~ e^{i\rho t} C_{n}(-t)
      \right] \\
    & = \dfrac{1}{2\pi} \left[ \wh{C}_{n}(i\rho) + \wh{C}_{-n}(-i\rho) \right]
  \end{split}
  \label{eq:wttCNrho-whCnirho}
\end{equation}
where we used
\begin{equation}
  \begin{split}
    C_{n}(-t)
    & = \avetau{\aveini{ Z_{-n}(-t+\tau)Z_{n}(\tau)}} \\
    & = \avetau{\aveini{ Z_{-n}(\tau)Z_{n}(t+\tau)}}
      = C_{-n}(t).
  \end{split}
\end{equation}

\begin{figure}[htbp]
  \centering
    \includegraphics[width=0.3\linewidth]{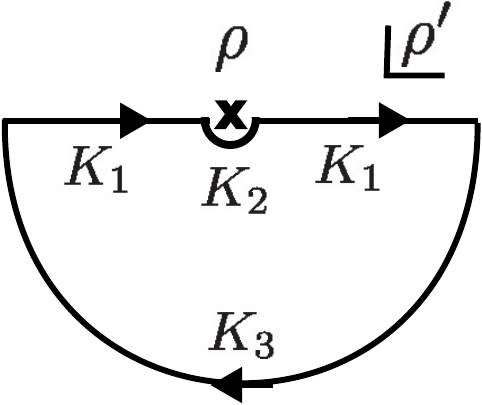}
  \caption{
    Integral contour $K=\cup_{i=1}^{3}K_{i}$, where $K_{1}$ is on the real axis.
    There is no singularity inside $K$.
    $K_{1}$ gives the principal value and $K_{2}$ the half residue.}
  \label{fig:KramersKronigContour}
\end{figure}

Second, substituting Eq.~\eqref{eq:wttCNrho-whCnirho} into Eq.~\eqref{eq:wttCnrho},
defining
\begin{equation}
  f(\rho) = -N \Lambda_{n}(i\rho) \Lambda_{-n}(-i\rho) \wh{C}_{n}(i\rho),
  \quad
  g(\rho) = \dfrac{1-\Lambda_{n}(i\rho)}{in\wt{\Gamma}_{n}},
\end{equation}
and noting $\Lambda_{-n}(s^{\ast})=[\Lambda_{n}(s)]^{\ast}$
and $\wh{C}_{-n}(-\rho)=[\wh{C}_{n}(i\rho)]^{\ast}$,
we have for $\rho\in\mathbb{R}$
\begin{equation}
  f(\rho) + [f(\rho)]^{\ast}
  = g(\rho) + [g(\rho)]^{\ast}
  \Longleftrightarrow
  {\rm Re}[f(\rho)] = {\rm Re}[g(\rho)].
\end{equation}
Here $[f(\rho)]^{\ast}$ is the complex conjugate of $f(\rho)$.
Let us consider the integrals
\begin{equation}
  F(\rho) = \dfrac{1}{\pi} \int_{K} \dfrac{f(\rho')}{\rho-\rho'} d\rho',
  \quad
  G(\rho) = \dfrac{1}{\pi} \int_{K} \dfrac{g(\rho')}{\rho-\rho'} d\rho',
\end{equation}
where the closed contour $K$ consists of $K_{1},K_{2}$, and $K_{3}$
(see Fig.~\ref{fig:KramersKronigContour}).
We have $F(\rho)=0$ since there is no singularity inside $K$.
Contribution from $K_{3}$ vanishes due to rapid decay of the integrands in $\rho'\to -i\infty$
and $F(\rho)=0$ implies
\begin{equation}
  \dfrac{1}{\pi} {\rm PV} \int_{-\infty}^{\infty} \dfrac{f(\rho')}{\rho-\rho'} d\rho'
  = i f(\rho),
  \label{eq:Kramers-Kronig}
\end{equation}
where PV represents the Cauchy principal value.
Therefore, the Kramers-Kronig relations
\begin{equation}
  {\rm Re}(f) = H[{\rm Im}(f)],
  \quad
  {\rm Im}(f) = -H[{\rm Re}(f)],
  \label{eq:Kramers-Kronig-2}
\end{equation}
hold, where $H[f]$ is the Hilbert transform of $f$ defined by the left-hand side of
Eq.~\eqref{eq:Kramers-Kronig}.
We can repeat the above discussion for $G(\rho)$ and we have
\begin{equation}
  {\rm Re}[f(\rho)] = {\rm Re}[g(\rho)]
  \Longleftrightarrow
  {\rm Im}[f(\rho)] = {\rm Im}[g(\rho)].
\end{equation}
Hence $f(\rho)=g(\rho)$ for $\rho\in\mathbb{R}$.

Third, the identity theorem states $f(\rho)=g(\rho)$ for $\rho\in\mathbb{C}$.
This equality gives the correlation function \eqref{eq:correlation}
by replacing $i\rho\in\mathbb{C}$ with $s\in\mathbb{C}$,
and the combination with the response \eqref{eq:response}
proves the fluctuation-response relation \eqref{eq:F-FRR}.

\subsection{Appendix B: Landau contour in the spectrum function}
\label{app_B}
The integral appearing in the spectrum function $\Lambda_{n}(s)$ is performed
along the Landau contour, which is a modified contour to avoid the singularity
at $\omega_{0}=i(s+n^{2}D)/n$. Hereafter we assume $n>0$. Note that the following explanation
is modified for $n<0$ by taking into account the position of $\omega_{0}$.

\begin{figure}[htbp]
  \centering
    \includegraphics[width=0.95\linewidth]{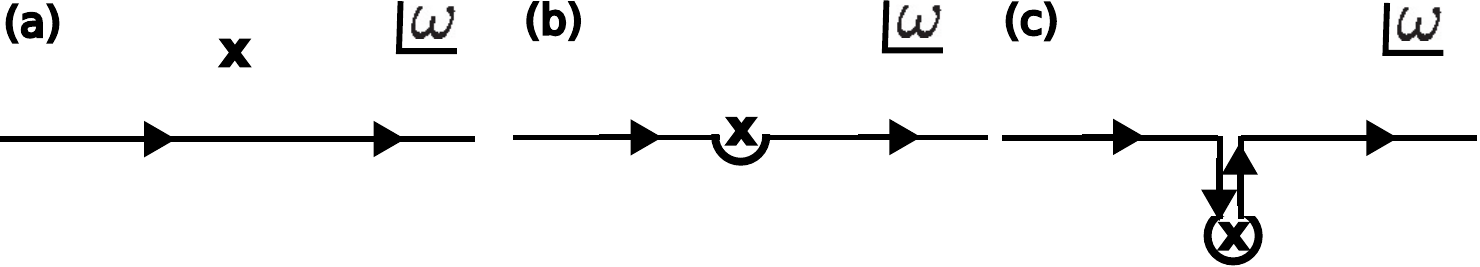}
  \caption{
    Landau contour on the complex $\omega$ plane for $n>0$.
    The horizontal line and the cross represent the real $\omega$ axis
    and the pole $\omega_{0}=i(s+n^{2}D)/n$ respectively.
    (a) ${\rm Re}(s)>-n^{2}D$. (b) ${\rm Re}(s)=-n^{2}D$. (c) ${\rm Re}(s)<-n^{2}D$.
  The pole moves from lower to upper for $n<0$
      and the sign of the residue becomes opposite.
  }
  \label{fig:LandauContour}
\end{figure}

The Laplace transform in $t$ requires ${\rm Re}(s)>0$ to ensure convergence
of the integral over $t$.
The pole $\omega_{0}$ is located in the upper-half complex $\omega$ plane
and the integral over the real $\omega$ axis is well-defined.
As ${\rm Re}(s)$ goes down, the pole $\omega_{0}$ arrives on the real axis
at ${\rm Re}(s)=-n^{2}D$, and the contour is modified to avoid it.
The modified contour contains the half residue.
As ${\rm Re}(s)$ is decreased further, the modification is continued,
and the residue contribution becomes full.
See Fig.~\ref{fig:LandauContour} for a graphical explanation.
The analytically continued integral in $\Lambda_{n}(s)$ is
\begin{equation}
  \begin{split}
    & \int_{L} \dfrac{ing(\omega)}{s+in\omega+n^{2}D} d\omega \\
    & = \left\{
      \begin{array}{ll}
        \displaystyle{ \int_{-\infty}^{\infty} \dfrac{g(\omega)}{\omega-i(s+n^{2}D)/n} d\omega } \\
        {\rm PV} \displaystyle{ \int_{-\infty}^{\infty} \dfrac{g(\omega)}{\omega-i(s+n^{2}D)/n} d\omega }
        + i\pi g(i(s+n^{2}D)/n) \\
        \displaystyle{ \int_{-\infty}^{\infty} \dfrac{g(\omega)}{\omega-i(s+n^{2}D)/n} d\omega }
        + i 2\pi g(i(s+n^{2}D)/n) \\
    \end{array}
  \right.
  \end{split}
\end{equation}
where ${\rm Re}(s)>-n^{2}D$, ${\rm Re}(s)=-n^{2}D$, and ${\rm Re}(s)<-n^{2}D$
from top to bottom.

\clearpage

\onecolumngrid

\begin{center}
    \textbf{\large Supplemental Material for ``Fluctuation–Response Relation in Finite-Size Noisy Coupled Phase Oscillators''}\\[5pt]
    Mrinal Sarkar$^1$ and Yoshiyuki Y. Yamaguchi$^2$ \\[3pt]
    \textit{$^1$Institut für Theoretische Physik, Universität Heidelberg, 69120 Heidelberg, Germany} \\[3pt]
    \textit{$^2$Graduate School of Informatics, Kyoto University, Kyoto 606-8501, Japan
} \\[3pt]
( Dated:\,\today)
\end{center}

\bigskip

\renewcommand*{\citenumfont}[1]{S#1}
\renewcommand*{\bibnumfmt}[1]{[S#1]}
\renewcommand{\theequation}{S\arabic{equation}}
\renewcommand{\thetable}{S\arabic{table}}
\renewcommand{\thefigure}{S\arabic{figure}}
\setcounter{equation}{0}
\setcounter{table}{0}
\setcounter{figure}{0}


\section{Details of computations of the correlation function $C_{n}(t)$}

In this section, we provide computations of the correlation function $C_{n}(t)$ in detail.
We start with the Fourier transform of $Z_{-n}(t)$ from $\zeta_{n}$,
\begin{equation}
  \wtt{Z}_{-n}(\rho)
  = \dfrac{1}{\Lambda_{n}(i\rho)}
  \sqrt{\dfrac{4\pi D}{N}}
  \int_{-\infty}^{\infty} \dfrac{in\sqrt{g(\omega)}}{i\rho+in\omega+n^{2}D}
  \wtt{\zeta}_{n}(\omega,\rho) d\omega.
\end{equation}
The inverse Fourier transform yields
\begin{equation}
  Z_{-n}(t) = \int_{-\infty}^{\infty} d\rho~ e^{i\rho t} \wtt{Z}_{-n}(\rho),
\end{equation}
and the correlation function $C_{n}(t)$ is
\begin{equation}
  \begin{split}
    C_{n}(t)
    & = \avetau{ \aveini{ Z_{-n}(t+\tau) Z_{n}(\tau) } }\\
    & = \avetau{
      \int_{-\infty}^{\infty} d\rho_{1} \int_{-\infty}^{\infty} d\rho_{2}~
      e^{i\rho_{1}(t+\tau)} e^{i\rho_{2}\tau}
      \aveini{\wtt{Z}_{-n}(\rho_{1}) \wtt{Z}_{n}(\rho_{2}) } }\\
    & = \avetau{ \dfrac{4\pi D}{N}
      \int_{-\infty}^{\infty} d\rho_{1} \int_{-\infty}^{\infty} d\rho_{2}~
      \dfrac{e^{i\rho_{1}(t+\tau)} e^{i\rho_{2}\tau}}{\Lambda_{n}(i\rho_{1})\Lambda_{-n}(i\rho_{2})}
      \int_{-\infty}^{\infty} d\omega_{1}
      \int_{-\infty}^{\infty} d\omega_{2}
      \dfrac{-(in)^{2}\sqrt{g(\omega_{1})g(\omega_{2})}
      \aveini{ \wtt{\zeta}_{n}(\omega_{1},\rho_{1}) \wtt{\zeta}_{-n}(\omega_{2},\rho_{2}) } }
      {(i\rho_{1}+in\omega_{1}+n^{2}D)(i\rho_{2}-in\omega_{2}+n^{2}D)} }.
  \end{split}
\end{equation}
The correlation function of $\zeta$ implies
\begin{equation}
  \begin{split}
    & \avenoise{ \wtt{\zeta}_{n}(\omega_{1},\rho_{1}) \wtt{\zeta}_{-m}(\omega_{2},\rho_{2}) }
    = \dfrac{1}{(2\pi)^{2}}
      \int_{-\infty}^{\infty} dt_{1} \int_{-\infty}^{\infty} dt_{2}
      e^{-i\rho_{1}t_{1}} e^{-i\rho_{2}t_{2}}
      \aveini{ \zeta_{n}(\omega_{1},t_{1}) \zeta_{-m}(\omega_{2},t_{2}) } \\
    & = \dfrac{1}{(2\pi)^{2}}
      \int_{-\infty}^{\infty} dt_{1} \int_{-\infty}^{\infty} dt_{2}
      e^{-i\rho_{1}t_{1}} e^{-i\rho_{2}t_{2}}
      \dfrac{1}{(2\pi)^{2}}
      \int_{-\pi}^{\pi} d\theta_{1} \int_{-\pi}^{\pi} d\theta_{2}
      e^{-in\theta_{1}} e^{im\theta_{2}}
      \aveini{ \zeta_{n}(\theta_{1},\omega_{1},t_{1}) \zeta_{-m}(\theta_{2},\omega_{2},t_{2}) } \\
    & = \dfrac{1}{(2\pi)^{4}}
      \int_{-\infty}^{\infty} dt_{1} \int_{-\infty}^{\infty} dt_{2}
      e^{-i\rho_{1}t_{1}} e^{-i\rho_{2}t_{2}}
      \int_{-\pi}^{\pi} d\theta_{1} \int_{-\pi}^{\pi} d\theta_{2}
      e^{-in\theta_{1}} e^{im\theta_{2}}
      \delta(\theta_{1}-\theta_{2}) \delta(\omega_{1}-\omega_{2}) \delta(t_{1}-t_{2})  \\
    & = \dfrac{1}{(2\pi)^{2}} \delta_{n,m} \delta(\omega_{1}-\omega_{2}) \delta(\rho_{1}+\rho_{2}). \\
  \end{split}
\end{equation}
Here we used
\begin{equation}
  \dfrac{1}{2\pi} \int_{-\pi}^{\pi} d\theta e^{-i(n-m)\theta} = \delta_{n,m},
  \quad
  \dfrac{1}{2\pi} \int_{-\infty}^{\infty} dt~ e^{-i\rho t} = \delta(\rho).
\end{equation}
Substituting this correlation into $C_{n}(t)$,
the $\tau$ dependence vanishes and we have
\begin{equation}
  \begin{split}
    C_{n}(t)
    = \dfrac{2n^{2}D}{2\pi N}
    \int_{-\infty}^{\infty} d\rho 
    \dfrac{e^{i\rho t}}{\Lambda_{n}(i\rho)\Lambda_{-n}(-i\rho)}
    \int_{-\infty}^{\infty} d\omega 
    \dfrac{g(\omega)}{(i\rho+in\omega+n^{2}D)(-i\rho-in\omega+n^{2}D)}.
  \end{split}
\end{equation}
Using the partial fraction decomposition
\begin{equation}
  \dfrac{1}{(i\rho+in\omega+n^{2}D)(-i\rho-in\omega+n^{2}D)}
  = \dfrac{1}{2n^{2}D} \left(
    \dfrac{1}{i\rho+in\omega+n^{2}D} +  \dfrac{1}{-i\rho-in\omega+n^{2}D} \right),
\end{equation}
we have
\begin{equation}
  C_{n}(t)
  = \dfrac{1}{2\pi N}  
  \int_{-\infty}^{\infty} d\rho
  \dfrac{e^{i\rho t}}{\Lambda_{n}(i\rho)\Lambda_{-n}(-i\rho)}
  \left(
    \dfrac{1-\Lambda_{n}(i\rho)}{-in\wt{\Gamma}_{n}}
    + \dfrac{1-\Lambda_{-n}(-i\rho)}{in\wt{\Gamma}_{-n}}
  \right)
\end{equation}
and the Fourier transform is expressed by Eq.~\eqref{eq:wttCnrho}.

\section{Fluctuation-response relation in the time domain}

The FRR is valid for any $n$:
\begin{equation}
  \wh{R}_{n}(s) = \wh{A}_{n}(s)\, \wh{C}_{n}^{N}(s),
  \label{eq:FRR-time-1}
\end{equation}
where $\wh{C}_{n}^{N}(s) = N\wh{C}_{n}(s)$ and
\begin{equation}
  \wh{A}_{n}(s) = -in\Lambda_{-n}(-s)\wh{H}_{n}(s)
  \label{eq:FRR-time-2}
\end{equation}
is theoretically computable. The FRR \eqref{eq:FRR-time-1} in the time
domain is
\begin{equation}
  R_{n}(t) = (A_{n} * C_{n}^{N})(t)
  = \int_{0}^{t} A_{n}(t-t')\, C_{n}^{N}(t')\, dt'.
  \label{eq:FRR-time-conv}
\end{equation}
This facilitates checking the FRR, since Eq.~\eqref{eq:FRR-time-1}
involves complex-valued functions on the complex $s$ plane.

We specialize to $n=-1$, since this is the index entering the response
$R_{-1}(t)$ tested throughout the main text. For
$H(\theta,t) = -h\,\Theta(t)\sin(\theta-\omega_{\rm ext}t)$, we have
\begin{equation}
  \wt{H}_{-1}(t) = -\dfrac{ih}{2}\Theta(t) e^{i\omega_{\rm ext}t},
  \qquad
  \wh{H}_{-1}(s) = -\dfrac{ih}{2}\, \dfrac{1}{s-i\omega_{\rm ext}}.
  \label{eq:FRR-time-H}
\end{equation}
Using Eq.~\eqref{eq:FRR-time-2} with $n=-1$,
\begin{equation}
  \wh{A}_{-1}(s) = i\,\Lambda_{1}(-s)\,\wh{H}_{-1}(s)
  = \dfrac{h}{2}\, \dfrac{\Lambda_{1}(-s)}{s-i\omega_{\rm ext}}.
  \label{eq:FRR-time-A-1}
\end{equation}
The inverse Laplace transform of $\wh{A}_{-1}(s)$ picks up the pole at
$s=i\omega_{\rm ext}$, yielding
\begin{equation}
  A_{-1}(t) = \dfrac{h}{2}\,\Lambda_{1}(-i\omega_{\rm ext})\, e^{i\omega_{\rm ext}t}.
  \label{eq:FRR-time-A-1-t}
\end{equation}
Thus, Eq.~\eqref{eq:FRR-time-conv} becomes, for $n=-1$,
\begin{equation}
  e^{-i\omega_{\rm ext}t} R_{-1}(t)
  = \dfrac{h}{2}\Lambda_{1}(-i\omega_{\rm ext})
  \int_{0}^{t} e^{-i\omega_{\rm ext}t'}\, C_{-1}^{N}(t')\, dt'.
  \label{eq:FRR-time-domain}
\end{equation}
This is the FRR in the time domain, and is precisely the quantity
compared to the directly measured response in
Figs.~\ref{fig:laplace_domain}, \ref{fig:fig2_timedomain}, and
\ref{fig:dk_validity_map}. $\Lambda_{1}(-i\omega_{\rm ext})$ is the
extra factor coming from the nonequilibrium nature of the finite-$N$
fluctuation.

\section{Numerical verification of the crossover of transient correlations}
\begin{figure}[htbp]
  \centering
  \includegraphics[width=0.8\linewidth]{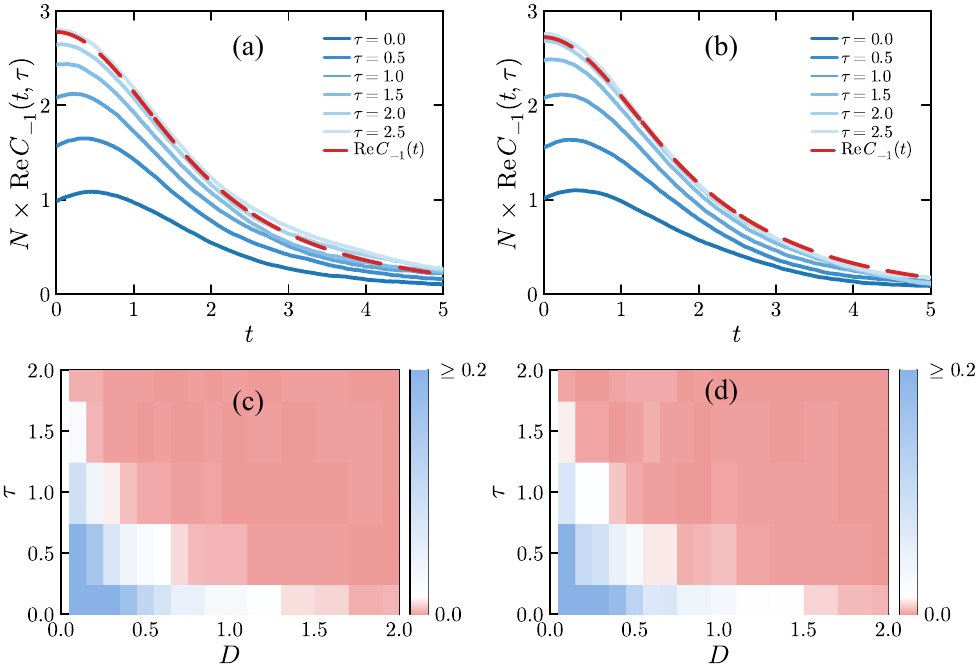}
  \caption{Numerical verification of the crossover of the transient correlation
$C_{-1}(t,\tau)$. Panels (a, b): $N\,\mathrm{Re}\,C_{-1}(t,\tau)$ for different reference times $\tau$
for $N=10^4$ and $10^5$, respectively, together with the stationary
correlation $N\,\mathrm{Re}\,C_{-1}(t)$ (red dashed line) at $K=1.0$ and $D=0.1$.
Panels (c, d): Corresponding $\Delta_{\rm Corr}(\tau,D)$ in the $(\tau,D)$ plane for
$N=10^4$ and $10^5$, respectively, showing a similar crossover timescale.
  }
  \label{fig_supp:corr_crossover}
\end{figure}

This section provides additional numerical support for the phenomena discussed in the main text and, in particular, validates the schematic picture in Fig.\,\ref{fig:Schematic}. To demonstrate the crossover and the correlations generated during relaxation, we compute the transient correlation function $C_{-1}(t,\tau)$ for different reference times $\tau$. Figures\,\ref{fig_supp:corr_crossover}(a) and \ref{fig_supp:corr_crossover}(b) show the resulting correlation functions for $N=10^4$ and $N=10^5$, respectively, at $K=1.0$ and $D=0.1$. In both cases, the transient correlation gradually approaches the stationary correlation as $\tau$ increases.

\begin{figure}[htbp]
  \centering
  \includegraphics[width=0.45\linewidth]{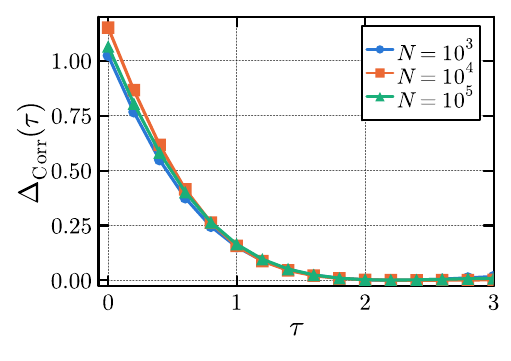}
  \caption{System-size dependence of the crossover in the weak-noise regime.
$\Delta_{\rm Corr}(\tau)$ for $K=1$ and $D=10^{-5}$ for
$N=10^3$, $10^4$, and $10^5$. The curves show a good collapse, indicating an approximately $N$-independent crossover timescale,
with increasing statistical fluctuations at larger $\tau$.}
  \label{fig_supp:corr_tau}
\end{figure}

To quantify this crossover, we define the mean-squared difference between the transient and stationary correlations,
\begin{equation}
\Delta_{\rm Corr}(\tau,D) = \frac{1}{T_{\rm obs}} \int_0^{T_{\rm obs}} \left| N\,\mathrm{Re}\,C_{-1}(t,\tau) - N\,\mathrm{Re}\,C_{-1}(t)
\right|^2 dt .
\end{equation}
The corresponding results are shown in Figs.\,\ref{fig_supp:corr_crossover}(c) and \ref{fig_supp:corr_crossover}(d) for $N=10^4$ and $N=10^5$, respectively. A clear crossover from large to small $\Delta_{\rm Corr}$ is evident in the $(\tau, D)$ plane. A nearly identical crossover time for $N=10^4$ and $10^5$ indicates that the crossover timescale is primarily set by the spectral function rather than by the system size. The absence of an explicit $N$-dependence is further illustrated in Fig.\,\ref{fig_supp:corr_tau}. For the weak-noise case $D=10^{-5}$ and $K=1$, the $\Delta_{\rm Corr}(\tau)$ curves for $N=10^3$, $10^4$, and $10^5$ collapse onto essentially the same curve, with increasing statistical fluctuations at larger $\tau$. This also supports the validity of the DK equation in the weak-noise regime. Note that Figs.\,\ref{fig_supp:corr_crossover}(c, d) numerically confirm the schematic diagram shown in Fig.\,\ref{fig:Schematic} in the main text.

\section{Statistical uncertainties}
\begin{figure}[htbp]
  \centering
  \includegraphics[width=0.90\linewidth]{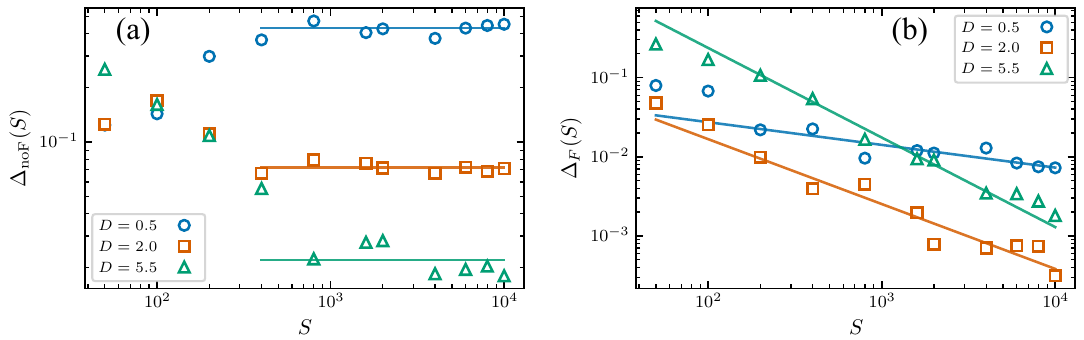}
  \caption{Dependence on the number of realizations $S$ used in the average, for $K = 1$ and three values of the noise strength $D$. (a) $\Delta_{\rm noF}(S)$ saturates to an approximately constant, $S$-independent plateau for each $D$, reflecting a residual difference between the two correlation functions that persists even with very large statistics. (b) $\Delta_{\rm F}(S)$ decreases with $S$; symbols are numerical data and solid lines are power-law fits $\Delta_{\rm F}(S) = a\,S^{-b}$, with fitted exponents $b \approx 0.28$, $0.8$, and $1.1$ for $D = 0.5$, $2.0$, and $5.5$, respectively. The decay of $\Delta_{\rm F}$ toward zero as $S \to \infty$, in contrast to the finite plateau of $\Delta_{\rm noF}$ in (a), confirms that the small nonzero values of $\Delta_{\rm F}$ mentioned in the main text are a finite-sampling effect.}
  \label{fig_supp:Delta_vs_S}
\end{figure}

This section provides numerical evidence that the small but nonzero values of $\Delta_{\rm F}$ observed in the regime where it appears to vanish in the main text (Fig.~\ref{fig:dk_validity_map}) are statistical uncertainties arising from finite-sampling fluctuations, rather than a genuine residual correlation. To this end, we fix $K = 1$ and consider three representative noise strengths, $D = 0.5$, $2.0$, and $5.5$, computing both $\Delta_{\rm noF}$ and $\Delta_{\rm F}$ as a function of the number of realizations $S$ over which each is averaged. As shown in Fig.~\ref{fig_supp:Delta_vs_S}(a), $\Delta_{\rm noF}(S)$ tends to saturate (within our numerical simulations) to an approximately constant, nonzero plateau for all three values of $D$. This indicates a residual difference between the two correlation functions that persists even with arbitrarily good statistics and is therefore not a finite-sampling fluctuation. By contrast, Fig.~\ref{fig_supp:Delta_vs_S}(b) shows that $\Delta_{\rm F}(S)$ decreases with $S$ for all three values of $D$. A power-law fit, $\Delta_{\rm F}(S) = a\,S^{-b}$, yields exponents $b \approx 0.29(5)$, $0.82(10)$, and $1.13(5)$ for $D = 0.5$, $2.0$, and $5.5$, respectively. Thus $\Delta_{\rm F}(S) \to 0$ as $S \to \infty$, unlike the finite plateau of $\Delta_{\rm noF}$, and the residual values are finite-sampling fluctuations of $\Delta_{\rm F}$ and vanish in the limit of very large statistics.  Although we have verified this explicitly only at these three representative values of $D$ and at $K=1$, we believe the same holds throughout the $(D,K)$ plane, except in the vicinity of the boundary (transition to synchronization) in Fig.~\ref{fig:dk_validity_map}.

\end{document}